# Effect of magnons on interfacial phonon drag in YIG/metal systems

Arati Prakash[1], Jack Brangham[1], Sarah J. Watzman[2], Fengyuan Yang[1], Joseph P. Heremans[1,2,3]

[1] Department of Physics, The Ohio State University, Columbus, Ohio 43210, USA

[2] Department of Mechanical Engineering, The Ohio State University, Columbus, Ohio 43210, USA

[3] Department of Materials Science and Engineering, The Ohio State University, Columbus, Ohio 43210, USA

**Abstract**

We examine substrate-to-film interfacial phonon drag on typical spin Seebeck heterostructures, in particular studying the effect of ferromagnetic magnons on the phonon-electron drag dynamics at the interface. We investigate with high precision the effect of magnons in the Pt|YIG heterostructure by designing a magnon drag thermocouple; a hybrid sample with both a Pt|YIG film and Pt|GGG interface accessible isothermally via a 6 nm Pt film patterned in a rectangular U shape with one arm on the 250 nm YIG film and the other on GGG. We measure the voltage between the isothermal ends of the U, while applying a temperature gradient parallel to the arms and perpendicular to the bottom connection. With a uniform applied temperature gradient, the Pt acts as a differential thermocouple. We conduct temperature-dependent longitudinal thermopower measurements on this sample. Results show that the YIG interface actually decreases the thermopower of the film, implying that magnons impede phonon drag. We repeat the experiment using metals with low spin Hall angles, Ag and Al, in place of Pt. We find that the phonon drag peak in thermopower is killed in samples where the metallic interface is with YIG. We also investigate magneto-thermopower and YIG film thickness dependence. These measurements confirm our findings that magnons impede the phonon-electron drag interaction at the metallic interface in these heterostructures.

**Introduction**

In this study, we focus on the longitudinal thermopower, $\alpha_{xxx}$, of a typical spin Seebeck heterostructure, i.e. Pt|YIG. In this configuration, a temperature bias is applied in the direction parallel to the direction that voltage is measured on the metal, i.e. along the direction of the interface. In comparison to an SSE measurement, in these studies we will essentially turn the sample on its side relative to the applied heat flux (see Figure 1).

Specifically, we are interested in studying the influence of magnons on the thermopower of the Pt|YIG heterostructure. We speculate that magnons at the Pt|YIG interface could exert a drag-like force on the electrons in the Pt, either directly or via an interaction mediated by phonons in YIG, and we endeavor to measure and discern this effect. In fact, when we aim to elucidate the parameter space of the SSE, which arises from magnon-phonon interactions, the question of drag between magnons, phonons and electrons in the Pt|YIG heterostructure is highly pertinent and is motivated by several contemporary studies, both experimental and theoretical.

Nonlocal drag (i.e. interfacial, between substrate and thin film) has been studied both in theory and experiment.[1,2] Ref. [1] examines nonlocal phonon-electron drag between an insulating sapphire substrate and a $Bi_2Te_3$ thin film, showing that the temperature dependent thermopower of the $Bi_2Te_3$ follows that of the thermal conductivity of the sapphire substrate, which implies that electronic transport in the film is enhanced by phononic transport in the sapphire substrate (phonon-electron drag). This demonstrates that substrate-to-thin-film phonon-electron drag can occur even when the two layers have dissimilar crystal structure. From this study, we would expect that there could be strong phonon-electron drag affecting the thermopower on either Pt|GGG or

Pt|YIG films. Furthermore, Ref. [2] predicts nonlocal magnon-magnon drag in a FM bilayer, arising from dipolar interactions across the interface.

Local drag effects (i.e. bulk, not interfacial) have also been studied: local magnon-electron drag (MED), or the advective transport of electrons dragged by magnons, has been examined in bulk metallic ferromagnets, where a thermal gradient drives magnons and electrons along with phonons.[3] While all metals contain drag and diffusive contributions to their thermopower, it was shown that the MED contribution actually dominated the thermopower in ferromagnetic metals (Fe, Co). The findings of thermopowers enhanced by magnon drag in Ref. [3] and of phonon drag dominating the thermopower even in dissimilar substrate-to-thin film heterostructures in Ref. [1], combined with the theoretical predictions in Ref. [2] that magnons can also participate in such nonlocal, interfacial drag effects, together inspire a look at magnon effects on the interfacial phonon drag on the Pt|YIG heterostructure.

Instead of an out of plane temperature gradient, which drives a nonlocal spin flux across the interface, we apply an in plane temperature gradient longitudinally as the driving force and examine magnon transport along this direction. We probe these dynamics by measurements of the longitudinal thermopower on the Pt film of Pt|YIG and similar heterostructures. As per Ref. [1], one would expect this thermopower as measured on the thin film to follow the thermal conductivity of the substrate (YIG or GGG) due to the interfacial phonon-electron drag. However, the question of interest here is what is the *effect* of magnons on this phonon electron drag?

To study this, we compare the thermopower of Pt films grown on ferrimagnetic YIG to that grown on paramagnetic GGG, where the only difference in substrate dynamics can be attributed purely to magnons in the YIG. In order to isolate the hypothetical drag contribution from the magnons in YIG into the adjacent Pt film, we design a thermocouple device using a hybrid sample with half Pt|GGG and half Pt|YIG(250nm)|GGG (see Figure 2). With a uniform applied temperature gradient, the Pt acts as a differential thermocouple. The effective voltage at the isothermal ends of the Pt provides a direct measure of the difference in thermopower of the two systems, which we attribute to magnon dynamics in YIG and their interactions at the Pt|YIG interface. The effective voltage at the isothermal ends of the Pt provides a direct measure of the difference in thermopower of the two systems, which we attribute to magnon dynamics in YIG and their interactions with phonons and electrons at the Pt|YIG interface.

Since the Pt|YIG (or Pt|YIG|GGG) heterostructure is the typical system used to examine spin thermal effects, we primarily focus on that system here. However, the large spin orbit coupling in Pt (which is precisely what makes it advantageous as the ISHE layer for SSE devices) could raise the question of contributions to the thermopower from SHE signals, especially in the presence of a magnetic field. To isolate any contribution from spin Hall effects, we repeat the experiment on similar heterostructures where the Pt metal is replaced by metals with rather low spin Hall angles. To examine magnonic effects on phonon electron drag (phonons in the YIG, electrons in the metallic thin film), we choose metals with relatively simple, clean Fermi surfaces as far as possible. We choose p-type Ag and n-type Al. When examining these metals in their thin film form to see how they interact with a substrate, we consider previous knowledge regarding the thermopower of these materials. Temperature dependent thermopower data for these bulk metals were measured decades back.[4,5] Although n-type, with a negative diffusion thermopower of -5 µV/K at 300 K, Pt exhibits a sign change in its thermopower around 200 K, with a *positive* phonon drag thermopower peak at 6 µV/K.[4] In contrast to this, Ag has a consistently positive (p-type) thermopower with a phonon drag peak near 1 µV/K and Al has a consistently negative (n-type)

thermopower with a phonon drag peak near -2 µV/K.[5] These values inform our interpretation of data in the context of relative strength of the thermopower measured here.

**Experiment**

In order to circumvent the influence of sample to sample variability on our measurements, we devise a hybrid heterostructure on which we can make a differential measurement $\alpha_{Pt|YIG}$ vs. $\alpha_{Pt|GGG}$ on one sample in situ. We name this hybrid heterostructure the magnon drag thermocouple (MDT). The MDT consists of 3 layers: a GGG substrate, a YIG film (250 nm) grown on *half* of the substrate with a gradually stepped edge at the longitudinal center fold of the sample, and a Pt film (6 nm) deposited across the entire structure then patterned into a squared-U shape with four corners (A, B, C, and D). In addition to the MDT for Pt, we construct an identical MDT for Ag (10 nm) and Al (6 nm). A list of MDT samples created can be found in Table 1.

All samples were measured for steady state, zero field thermopower, $\alpha_{xxx}$, in the static-heat-sink configuration using a Quantum Design PPMS as in Ref. [3], with thermometry and gold plated copper leads attached to the back face of each substrate. Voltage probes to measure thermopower were attached via small (~25 µm) Ag epoxy contacts placed directly on the thin film. With a temperature gradient applied longitudinally, the two arms (AB and DC) comprise the thermocouple on which a differential voltage ($V_{ad}$) can be measured. The hybrid heterostructure acts effectively as a thermocouple for the Pt interface: because both sides of the bottom bar (B and C) are isothermal with an applied longitudinal temperature gradient, any voltage measured across AD would be due to a difference in the voltage drop across arm AB vs arm DC, i.e., a difference in the interfacial thermopower depending on whether or not YIG is present.

One can reasonably ask the question of the influence of the bottom bar of the U on the signal. Depending on the direction of the applied magnetic field, this would correspond to a Nernst-like configuration or transverse spin Seebeck effect (see Figure 3). Upon the addition of an applied magnetic field, such a point becomes relevant. This question is simply addressed by a measurement across the bar, revealing little-to-none signal on the $V_{bc}$ channel, a result which could also have been predicted noting that the Nernst effect is ten times smaller (often 1 in 2000) than the Seebeck effect in metals.[6]

| Sample | Film Deposition |
|---|---|
| Pt(6 nm)|half YIG(250nm)|GGG | U |
| Al(6 nm)|half YIG(250 nm)|GGG | U |
| Ag(6 nm)|half YIG(250 nm)|GGG | U |
| Pt(6 nm)|GGG (control) | U (no half-YIG) |

Table 1. Directory of magnon drag thermocouple (MDT) samples.

**Results**

In order to characterize our Pt films, we measure temperature dependence of the resistivity using the standard AC Transport option in the PPMS. We measure in zero field and at 7 Tesla applied magnetic field; results show no anomalies and the resistance behaves as expected (see Figure 4). We also measure the thermal conductivity of every substrate used in this study, in situ with the Seebeck measurements. An example of thermal conductivity of GGG is shown in Figure 5. These measurements help to keep track of sample quality and check for induced defects as the study goes on. As is characteristic for phonon thermal conductivity, there is a low temperature drop off, where the density of carriers decreases as $T^3$. The high temperature drop is attributed to intrinsic phonon-phonon Umklapp scattering, and the low temperature drop is attributed to phonon-boundary scattering.[7] The intermediate temperature range is where phonon transport is maximum and where the phonon drag peak in the thermopowers is expected to be maximal.

To test of the validity of the MDT, we measure a control sample of a Pt U deposited on a GGG substrate, with no half-YIG film. Although in principle there should clearly be no signal on $V_{ad}$ of such a sample, this measurement demonstrates the validity of the assumption of the U as a reliable thermocouple in practice. Measurements confirm there is no spurious signal from the bottom bar, and that $V_{ad}$ is isovoltaic in the absence of YIG. The low temperature measurements from this control reveal the baseline noise level of the experiment, on the order of 1 µV/K below 6 K.

Seebeck measurements from the MDT in differential mode ($V_{ad}$) are shown in Figure 6. The data show a temperature dependence roughly following that of the thermal conductivity of the substrate GGG and a positive peak near 8 µV/K. It is worth noting carefully that measurements of the MDT in differential mode actually give the thermopower of Pt on YIG subtracted from the thermopower of Pt on GGG, considering the cold side to be the positive voltage terminal, as is conventional in Seebeck measurements. This means that the positive signal on the Pt MDT in Figure 6 implies that the magnons *lower* the thermopower of the Pt on YIG relative to the Pt on GGG. This result is surprising, as it implies that magnons at the interface may actually be suppressing the drag effects across the Pt interface.

Next, we examine the differential thermopower of the Ag and Al MDTs. Two observations are immediately evident: 1) the thermopower magnitudes of each of the films exceed those of their bulk counterparts and 2) their temperature dependence roughly follows thermal conductivity of substrate, implying phonon electron drag, substrate to film. A positive low temperature peak is around 4 µV/K on the Ag sample, and a negative thermopower with a peak is around -14 µV/K in the Al. Given that Ag is p type and Al is n type, we find that the polarity of the effect matches that of the sign of the carrier in the metal. This verifies that the measured thermopower voltage is not related to spin Hall physics, but rather electron drag by phonons. Thus, these results are consistent with the results on the Pt sample; the YIG yields a lower signal than the GGG interface, implying that the magnons interfere with phonon drag.

As a follow up to this observation, we conduct magnetic field-dependent measurements of the thermopower on the Pt and Ag MDTs, as shown in Figure 7. Applied magnetic fields are expected to suppress or "freeze out" magnon dynamics,[8] with a more pronounced effect at low temperatures.[9] At lowest temperatures, where one would expect the effect of the magnetic field to be strongest (6 to 9 K) the interfacial thermopower itself is difficult to resolve, so that a field dependent study is difficult to obtain. At moderately low temperatures where signal is strongest, near the phonon drag peak in thermopower, this magnetic field effect is measurable. As the magnetic field is increased from 0 to 9 Tesla at 10 K, we observe a decrease in thermopower on both Pt and Ag films. A decrease in thermopower implies a decrease in signal as magnon dynamics are suppressed at large magnetic fields so that both YIG and GGG exert the same amount of drag on the Pt. This effect is more pronounced at low temperatures, but where signal is still well resolvable from the noise floor, which becomes difficult below around 7 K. By contrast, we also measure the magneto-Seebeck coefficient of a Pt|YIG(250) sample. Here, we observe an *increase* in thermopower as magnetic field is increased from 0 to 9 Tesla below 30 K, as shown in Figure 8. An increase in thermopower implies a *recovery* of signal as magnon dynamics are suppressed out at large magnetic fields; this result is consistent with the results in Refs. [8,9] and differential measurements from the Pt MDT, supporting the conclusion that magnons interfere with phonon drag in these heterostructures.

Having isolated that there is a magnonic impedance to the phonon drag effect, we explore the length scale of this effect. In particular, as we decrease YIG thickness, there should be less magnons available, so that the Pt|YIG thermopower should increase and ultimately for very thin YIG, match that of Pt|GGG. We now measure a series of Pt|YIG samples with YIG of varied thickness (bulk, 1 μm, 250 nm, 100 nm, 40 nm). The bulk sample behaves much like the 250 nm YIG sample. At 100 nm, the signal increases, and at 40 nm, the phonon drag peak in the Pt|YIG thermopower is nearly equivalent to that of the Pt|GGG. In the 1 μm samples, the phonon drag peak is killed altogether, but the diffusion thermopower (high temperatures) equilibrates for all samples above around 100 K.

The implications of these data are summarized as follows. Substrate-to-thin film phonon electron drag is observed on Pt|YIG and Pt|GGG with equal magnitudes at high temperatures. The phonon drag peak in thermopower is significantly attenuated in the metal when YIG is present, so that this attenuation is attributed to magnons in the YIG. The thicker the YIG film, the larger the magnon scattering volume, which effectively acts as a barrier for the phonons which otherwise would drag electrons in the neighboring conducting film. At the smallest YIG thicknesses, we recover results very similar in magnitude to the signals on GGG. This length scale dependence on YIG thickness complements the identification of the magnon energy relaxation length from Ref. [10]. In fact the magnon-to-phonon energy relaxation, or a difference between magnon and phonon temperatures, could reasonably affect scattering rates between magnons and phonons and consequently interrupt the phonon-electron drag effects that occur in the absence of magnons.

With an applied magnetic field at temperatures below 30 K, we observe a recovery of the phonon-electron drag as magnon dynamics are partially suppressed. The effect manifests as an increase or recovery in signal on the isolated YIG system, and a decrease in differential signal on the MDT. These observations are consistent with the magnetic field-dependent "freeze out" of magnon dynamics established in Refs. [8,9]: as magnons are suppressed with high magnetic fields, the Pt|YIG interface behaves more closely like the Pt|GGG.

This series of measurements support our conclusion that magnons in fact interfere with the phonon-electron drag interaction at the metallic interface in these heterostructures. Further work should focus on developing a quantitative theoretical model for such an effect, accounting for scattering rates of magnons with phonons and electrons in the YIG and at the Pt interface.

**Acknowledgements**

Funding for this work was provided by the OSU Center for Emergent Materials, an NSF MRSEC, Grant DMR-1420451 and the U.S. Department of Energy, Office of Science, Basic Energy Sciences, Grant No. DE-SC0001304.

**Figures**

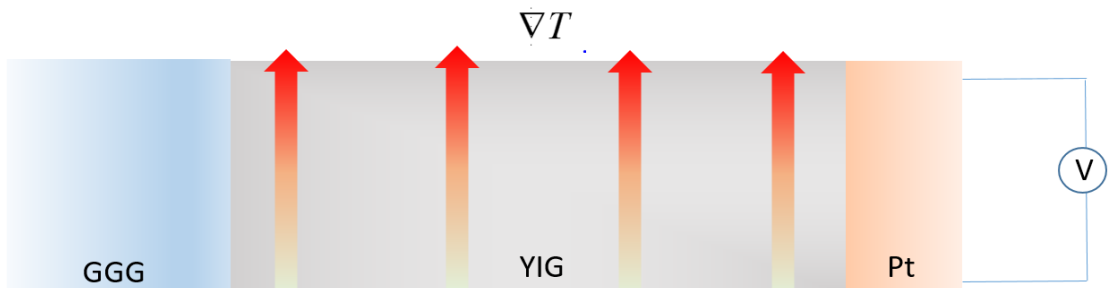

Figure 1. Schematic of longitudinal thermopower measurement, where voltage is measured in the direction of the temperature bias, along the direction of the interface on Pt|YIG heterostructure.

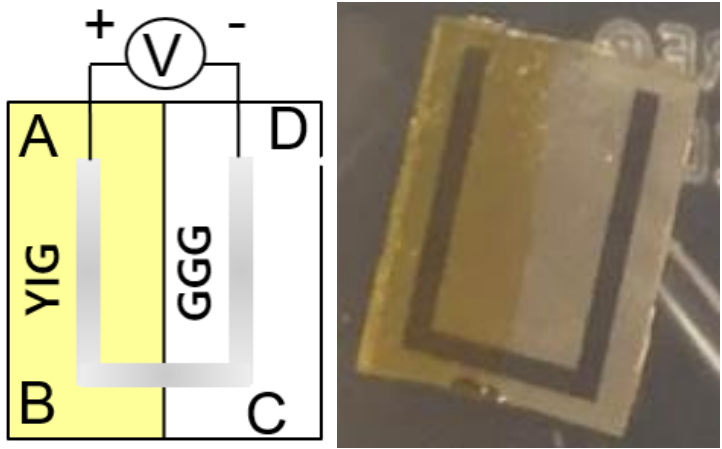

Figure 2. Schematic of magnon drag thermocouple with measured voltage in differential mode (Vad) (left), and photograph of actual Pt|YIG(/GGG) U shaped sample (right).

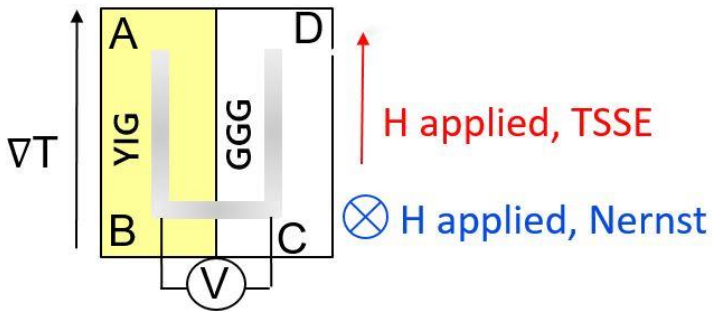

Figure 3. The direction of the applied magnetic field yields either a Nernst configuration (blue) or a transverse spin Seebeck configuration (red) on the bottom bar of the MDT.

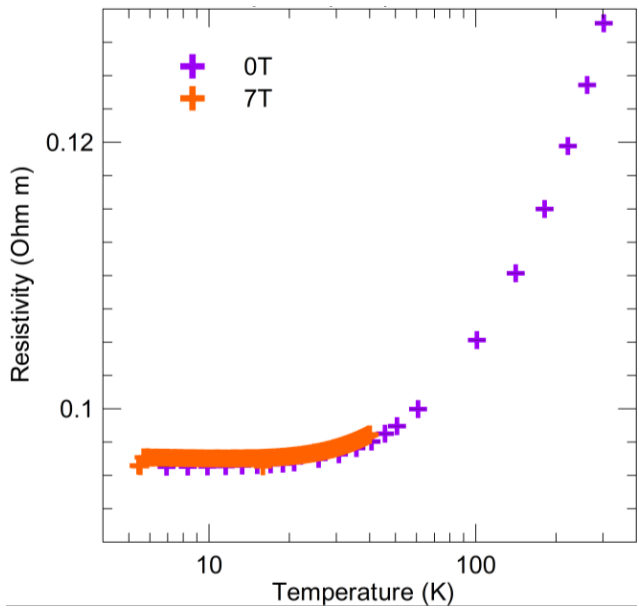

Figure 4. Temperature dependent resistivity of Pt on GGG in 7 T field (orange) and in zero-field (purple) shows no anomalous features and serves as an experimental check of the Pt film.

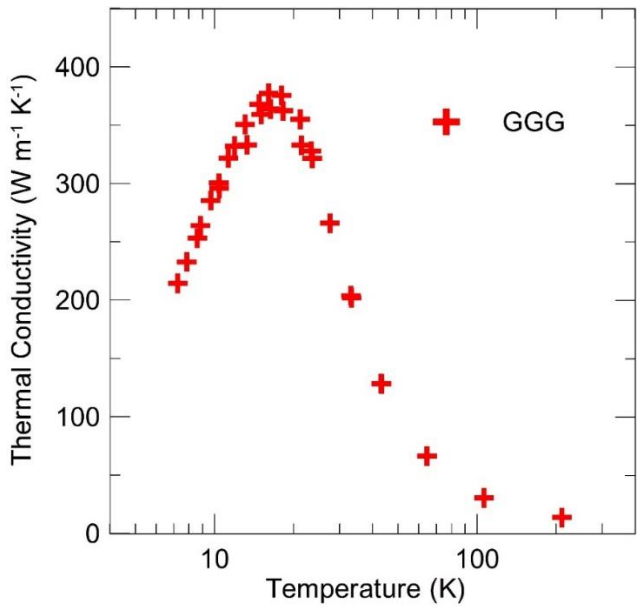

Figure 5. Temperature dependent thermal conductivity of bulk single crystal GGG substrate.

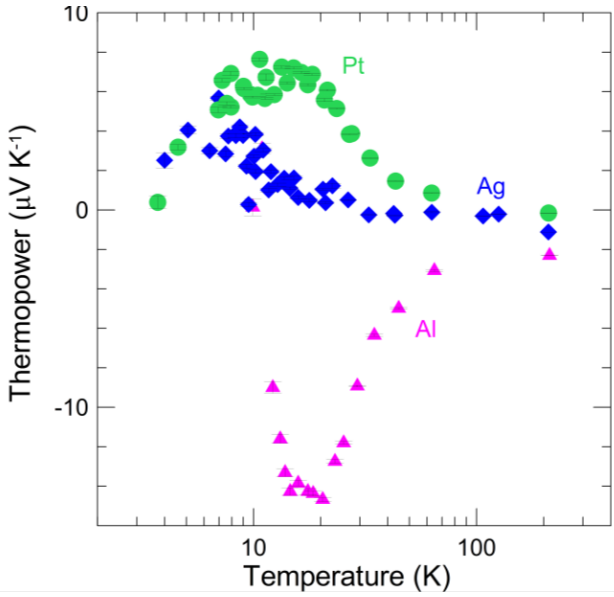

Figure 6. Temperature dependent thermopower of Pt, Ag, and Al magnon drag thermocouples.

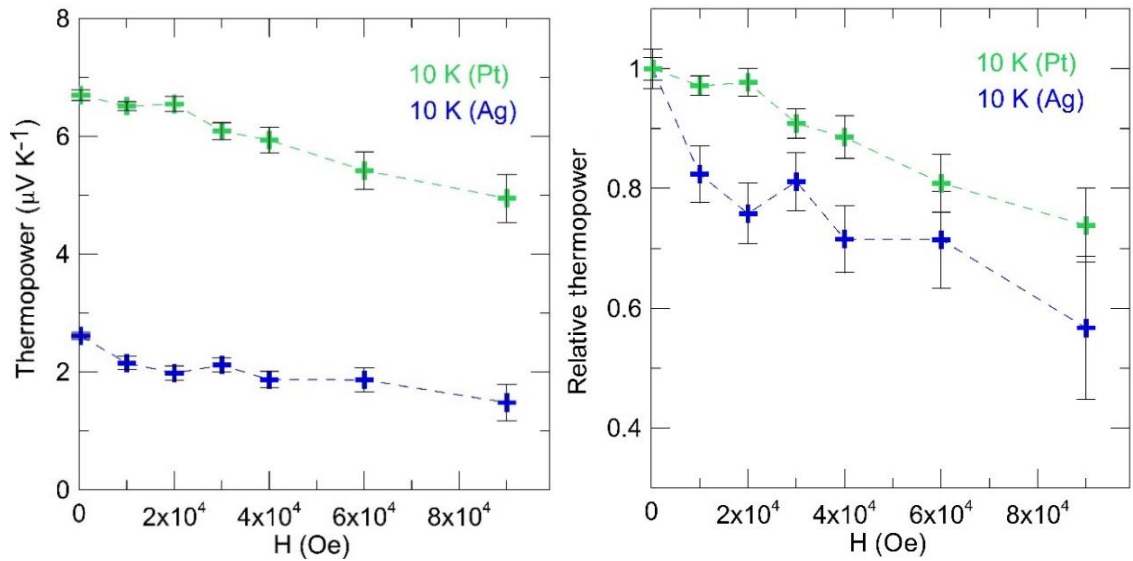

Figure 7. Magnetic field dependence of thermopower at various temperatures indicated on the graphs in the Pt and Ag magnon drag thermocouples in differential mode (left). The relative signal can be seen as a relative decrease from the zero field value (right).

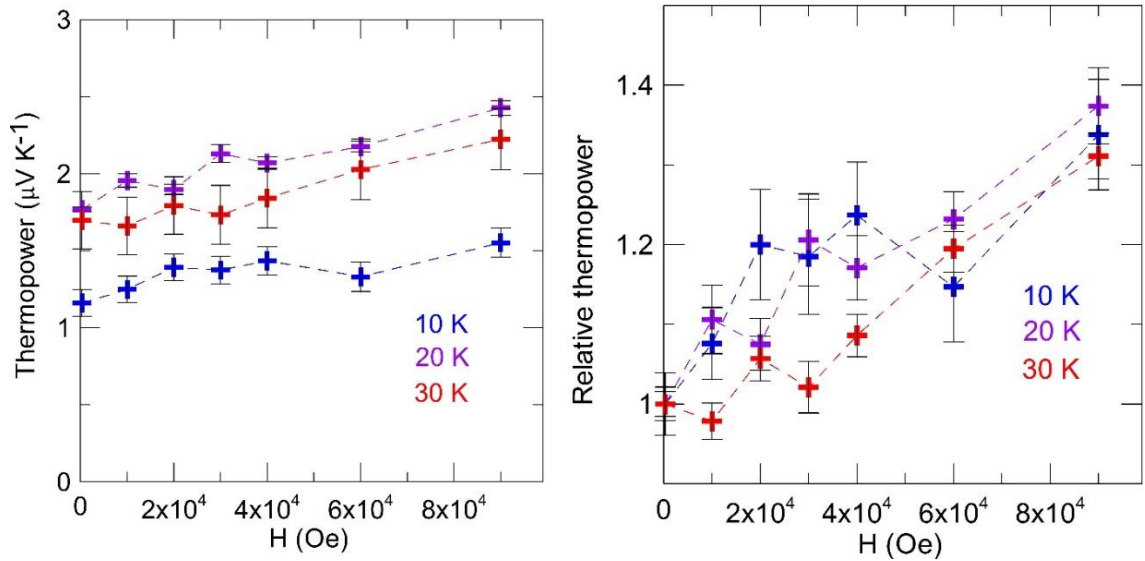

Figure 8. Magnetic field dependence of thermopower, at various base temperatures indicated on the graphs for the Pt|YIG interface (left). The strength of this effect can be seen as a relative increase from zero-field thermopower (right).

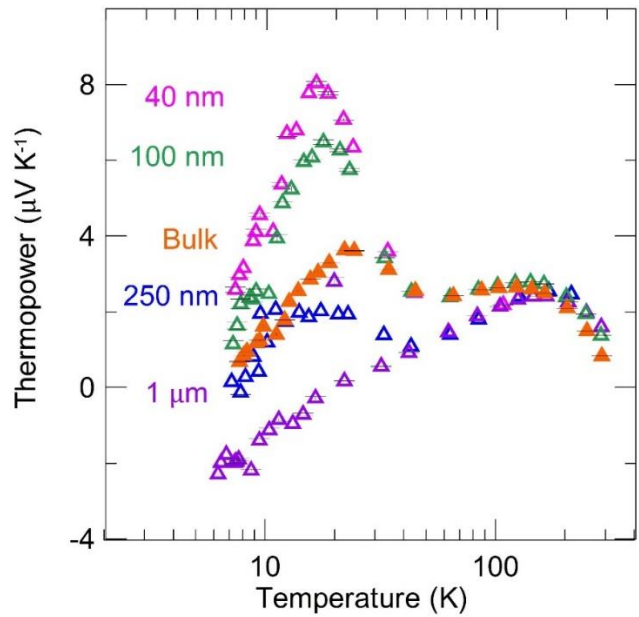

Figure 9. Temperature dependence of interfacial Pt|YIG thermopower for various YIG thicknesses as shown on the graph.